# Scattering and extinction by spherical particles immersed in an absorbing host medium

Michail I. Mishchenko[a,*] and Janna M. Dlugach[b]

[a] *NASA Goddard Institute for Space Studies, 2880 Broadway, New York, NY 10025, USA*
[b] *Main Astronomical Observatory of the National Academy of Sciences of Ukraine, 27 Zabolotny Str., 03680, Kyiv, Ukraine*

*Corresponding author: michael.i.mishchenko@nasa.gov

ABSTRACT

Many applications of electromagnetic scattering involve particles immersed in an absorbing rather than lossless medium, thereby making the conventional scattering theory potentially inapplicable. To analyze this issue quantitatively, we employ the FORTRAN program developed recently on the basis of the first-principles electromagnetic theory to study far-field scattering by spherical particles embedded in an absorbing infinite host medium. We further examine the phenomenon of negative extinction identified recently for monodisperse spheres and uncover additional evidence in favor of its interference origin. We identify the main effects of increasing the width of the size distribution on the ensemble-averaged extinction efficiency factor and show that negative extinction can be eradicated by averaging over a very narrow size distribution. We also analyze, for the first time, the effects of absorption inside the host medium and ensemble averaging on the phase function and other elements of the Stokes scattering matrix. It is shown in particular that increasing absorption significantly suppresses the interference structure and can result in a dramatic expansion of the areas of positive polarization. Furthermore, the phase functions computed for larger effective size parameters can develop a very deep minimum at side-scattering angles bracketed by a strong diffraction peak in the forward direction and a pronounced backscattering maximum.

## 1. Introduction

Many natural as well as artificial environments feature discrete scattering particles embedded in an absorbing rather than lossless host, which can make traditional theoretical treatments of electromagnetic scattering [1,2] potentially inapplicable [3–21]. Important geophysical examples of lossy host media are ice and water bodies at infrared wavelengths as well as strongly absorbing gaseous atmospheres, while the range of examples of artificial absorbing hosts is virtually unlimited.

The recent computer implementation [22] of a first-principles electromagnetic scattering theory [23] has made possible a systematic quantitative study of relevant far-field single-scattering optical observables for homogeneous spherical particles immersed in an absorbing unbounded medium. The focus of a subsequent initial numerical analysis [24] was on the extinction efficiency factor

$$Q_{\text{ext}} = \frac{C_{\text{ext}}}{\pi R^2} \qquad (1)$$

of a *monodisperse* sphere, where $R$ is the radius of the particle and $C_{ext}$ is its extinction cross section. Specifically, it was examined how $Q_{ext}$ depends on the particle size parameter

$$x = \frac{2\pi R}{\lambda} \qquad (2)$$

and the complex refractive indices of the particle, $m_2 = m_2' + im_2''$, and the host medium, $m_1 = m_1' + im_1''$, where, $\lambda$ is the vacuum wavelength and $i = (-1)^{1/2}$. It has been demonstrated that if $m_2'' = 0$ then the suppressing effect of increasing $m_1''$ on the ripple structure of $Q_{ext}$ as a function of $x$ is analogous to the well-known effect of increasing $m_2''$ of a particle embedded in a nonabsorbing host ($m_1'' = 0$). However, the effect of increasing $m_1''$ on the interference structure of the extinction efficiency curves is dramatically different from that of increasing $m_2''$, so that sufficiently large absorption inside the host medium can cause *negative* values of $Q_{ext}$ for a particle made of a lossless material. The simple physical explanation of the phenomenon of negative extinction suggested in Ref. [24] is consistent with the interpretation of the interference structure as being the result of interference of the electromagnetic fields directly transmitted and diffracted by the particle [1,25,26].

This paper is a natural outgrowth of Ref. [24] and has three primary objectives. The first one is to further analyze the phenomenon of negative extinction and look for additional evidence in favor of its interference origin. The second objective stems from the fact that in typical practical applications one has to deal with ensembles of particles distributed over sizes. Therefore, we will compare extinction efficiency curves computed for *polydisperse* spheres with those of monodisperse particles and identify the main effects of increasing the width of the size distribution on the ensemble-averaged $Q_{ext}$. The third objective is to study the effects of absorption inside the host medium as well as ensemble averaging on such key far-field optical observables as the phase function and other elements of the Stokes scattering matrix.

## 2. Negative extinction

The remarkable phenomenon of negative extinction identified in Ref. [24] is illustrated by Fig. 1 depicting the extinction efficiency factor $Q_{ext}$ for an

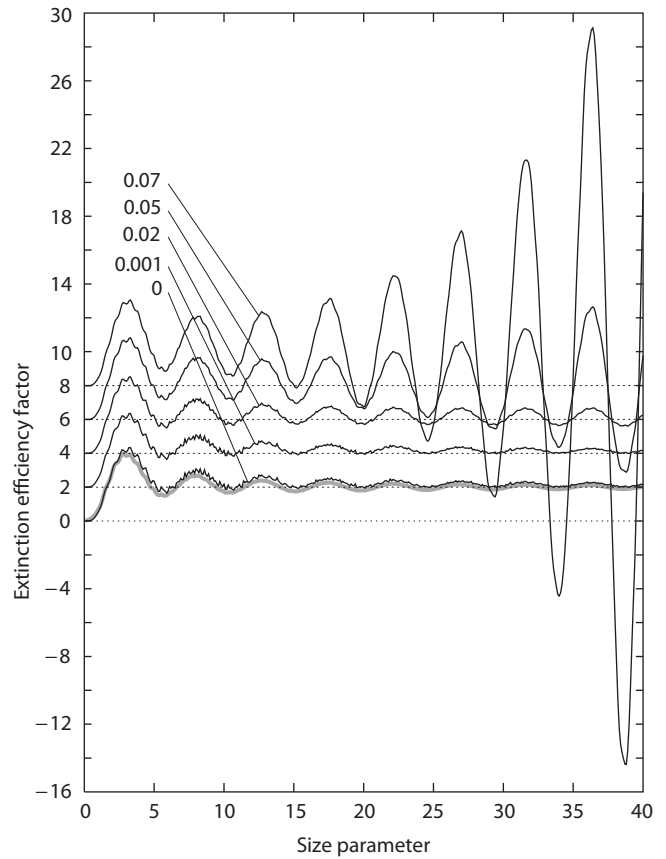

**Fig. 1.** Thin black curves: the monodisperse extinction efficiency factor versus the size parameter for $m_1' = 1.33$ and $m_2 = 2$. The five extinction curves are labeled by the respective values of $m_1''$. The vertical scale applies to the curve for $m_1'' = 0$, the other curves being successively displaced upward by 2. The thick gray curve shows the result of using Eq. (3). The results are displayed with a size-parameter step size of $\Delta x = 0.1$.

isolated spherical particle as a function of the particle size parameter $x$. It is seen that all five thin black curves are dominated by quasi-regular oscillations forming the so-called interference structure [1]. When the host medium is nonabsorbing ($m_1'' = 0$), the amplitude of these oscillations decreases with increasing $x$, causing $Q_{ext}$ to approach its asymptotic value 2. This behavior is called the extinction paradox [1,27,28]. If, however, $m_1''$ exceeds a certain threshold, the amplitude of the oscillations increases rather than decreases with increasing $x$, ultimately causing negative values of the extinction efficiency factor. Note that this frequency-domain phenomenon is fundamentally different from negative extinction

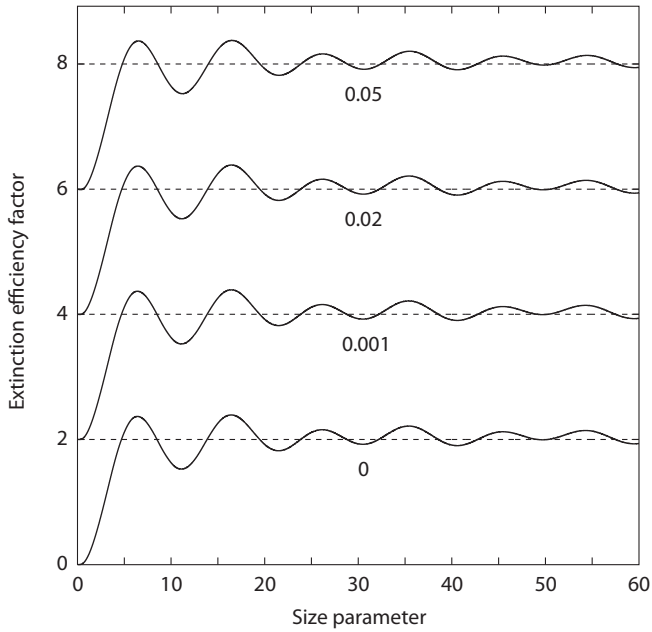 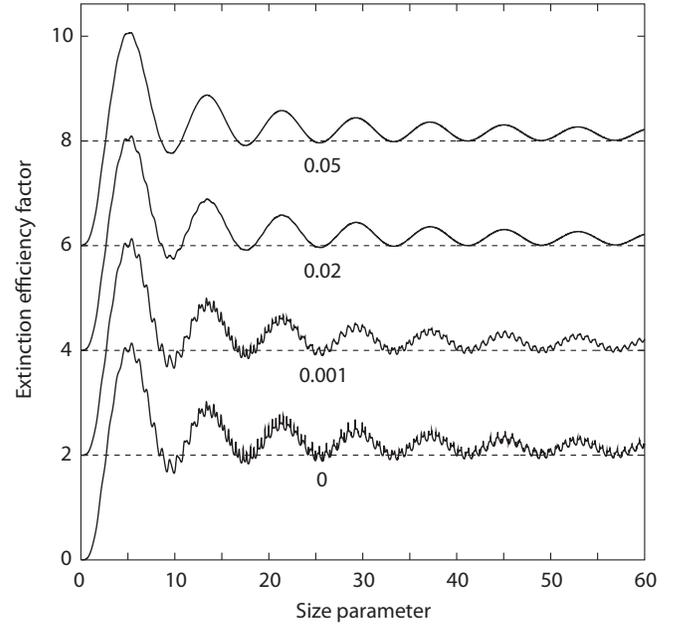

**Fig. 2.** The monodisperse extinction efficiency factor versus the size parameter for $m'_1 = 1.33$ and $m'_2 = 1$. The curves are labeled by the corresponding values of $m''_1 = m''_2$. The vertical scale applies to the curve for $m''_1 = m''_2 = 0$, the other curves being successively displaced upward by 2. The results are displayed with a size-parameter step size of $\Delta x = 0.01$.

**Fig. 3.** The monodisperse extinction efficiency factor versus the size parameter for $m'_1 = 1$ and $m'_2 = 1.4$. The curves are labeled by the value of $m''_1 = m''_2$. The vertical scale applies to the curve for $m''_1 = m''_2 = 0$, the other curves being successively displaced upward by 2. The results are displayed with a size-parameter step size of $\Delta x = 0.01$.

potentially exhibited by active particles [29–32] or by transient electromagnetic scattering [33].

The qualitative explanation of negative extinction suggested in Ref. [24] is based on the premise that if there were no differential attenuation between the electromagnetic fields "directly transmitted" and "diffracted" by the particle then the interference structure of $Q_{\text{ext}}$ would largely be similar to that exhibited by the $m''_1 = 0$ extinction curve. If, however, the host medium is absorbing and $m''_2 = 0$ then the directly transmitted field is no longer subject to the exponential attenuation over the path length given by the particle diameter. This causes an exponentially growing amplitude of the interference oscillations with increasing $x$.

This simple qualitative explanation is corroborated by the thick gray curve in Fig. 1 showing the result of replacing $Q_{\text{ext}}$ computed for $m''_1 = 0.07$ by

$$Q'_{\text{ext}} = (Q_{\text{ext}} - 2)\exp(-2k''R) + 2. \qquad (3)$$

where $k''_1 = 2\pi m''/\lambda$. It is seen indeed that this curve replicates closely the thin black curve computed for $m''_1 = 0$, which is in agreement with analogous results of Ref. [24].

To further corroborate the interference origin of negative extinction, we have performed additional computations of the extinction efficiency factor for scenarios wherein both $m''_1$ and $m''_2$ are non-zero and are equal to each other. The results of these computations are summarized in Figs. 2 and 3. The combination of $m'_1 = 1.33$ and $m'_2 = 1$ in Fig. 2 can be thought of as representing an air bubble in water or water ice, while the combination of $m'_1 = 1$ and $m'_2 = 1.4$ in Fig. 3 can represent a liquid or solid particle suspended in an absorbing gas.

It is clearly seen that irrespective of the degree of absorption in the host medium and in the particle, $Q_{\text{ext}}$ is always positive. The only obvious effect of increasing absorption is to weaken and ultimately eradicate the fine ripple structure in Fig. 3, while the degree of similarity of the curves in Fig. 2 is quite remarkable. Consistent with the explanation in Ref. [24], the ripple structure consisting of quasi-randomly

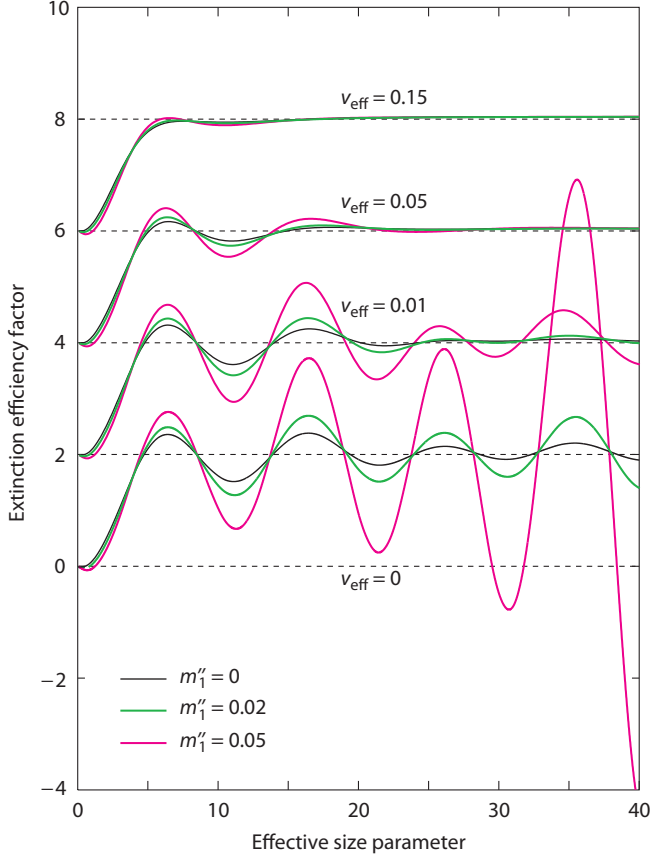

**Fig. 4.** The polydisperse extinction efficiency factor versus the effective size parameter for $m'_1 = 1.33$, $m'_2 = 1$, and $m''_1 = 0$, 0.02, and 0.05. Each set of three extinction curves is labeled by the corresponding value of $v_{\text{eff}}$. The vertical scale applies to the curve for $v_{\text{eff}} = 0$, the other curves being successively displaced upward by 2. The results are displayed with a size-parameter step size of $\Delta x_{\text{eff}} = 0.01$.

---

positioned sharp morphology-dependent resonances (MRDs) is completely absent in Fig. 2 even when $m''_1 = m''_2 = 0$ because total internal reflection is impossible if $m_2/m_1 < 1$.

Thus the results depicted in Figs. 2 and 3 substantiate quite convincingly the main premise of the interference explanation of the phenomenon of negative absorption according to which negative $Q_{\text{ext}}$ values are caused by a sufficiently strong *differential* attenuation between the fields transmitted and diffracted by the particle [24].

### 3. Effects of polydispersion

In the majority of practical applications, scattering particles are *polydisperse* rather than have exactly the same size. Therefore, the quasi-regular oscillatory behavior of the interference structure computed for monodisperse particles makes it interesting to analyze whether the phenomenon of negative extinction can survive averaging over an increasingly wide size distribution. To this end, we average the extinction cross section $C_{\text{ext}}$ over the conventional gamma distribution of particle radii given by [34,35]

$$n(R) = \text{constant} \times R^{(1-3b)/b} \exp\left(-\frac{R}{ab}\right), \quad b \in (0, 0.5), \tag{4}$$

where the constant is chosen such that the size distribution satisfies the standard normalization

$$\int_0^\infty dR\, n(R) = 1. \tag{5}$$

Two canonical characteristics of the size distribution [34,35] are the effective radius $r_{\text{eff}}$ and effective variance $v_{\text{eff}}$ defined, respectively, by

$$r_{\text{eff}} = \frac{1}{\langle G \rangle_R} \int_0^\infty dR\, n(R) R \pi R^2 \tag{6}$$

and

$$v_{\text{eff}} = \frac{1}{\langle G \rangle_R r_{\text{eff}}^2} \int_0^\infty dR\, n(R)(R - r_{\text{eff}})^2 \pi R^2 \equiv b, \tag{7}$$

where

$$\langle G \rangle_R = \int_0^\infty dR\, n(R) \pi R^2 \tag{8}$$

is the average area of the geometric projection per particle. According to Eq. (6), $r_{\text{eff}}$ is the projected-area-weighted mean radius, whereas the dimensionless effective variance can be thought of as defining a relative measure of the width of the size distribution. The distribution with $v_{\text{eff}} = 0$ corresponds to monodisperse particles. The polydisperse extinction efficiency factor is now defined according to

$$Q_{\text{ext}} = \frac{\langle C_{\text{ext}} \rangle_R}{\langle G \rangle_R} = \frac{1}{\langle G \rangle_R} \int_0^\infty dR\, n(R) C_{\text{ext}}(R). \tag{9}$$

To maximally reveal the effect of ensemble averaging on negative extinction, the results depicted in Figs. 4 and 5 are computed by assuming that the particle material is lossless. In this case the ensemble-averaged extinction efficiency factor (9) is plotted as a function of the effective size parameter defined by

$$x_{\text{eff}} = \frac{2\pi r_{\text{eff}}}{\lambda}. \qquad (10)$$

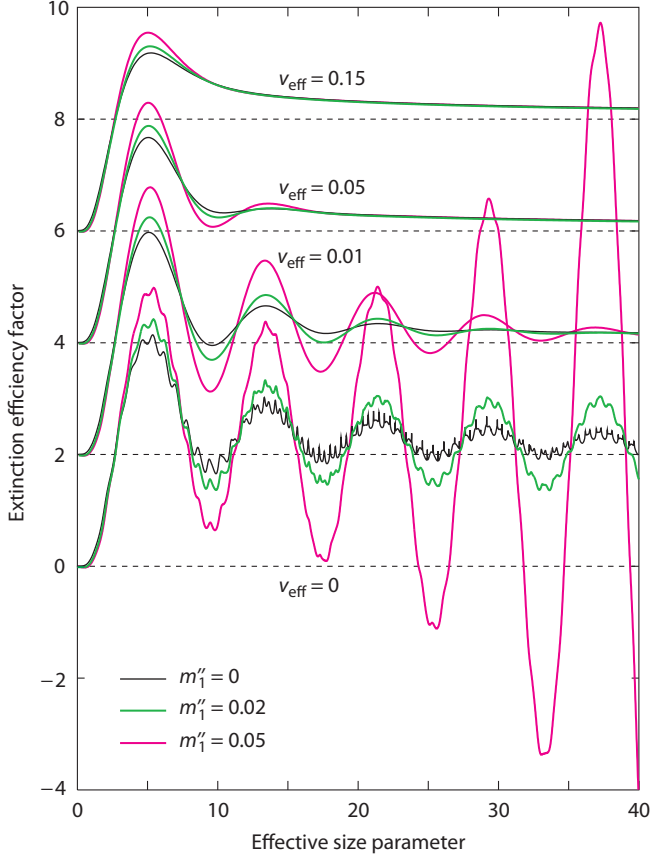

**Fig. 5.** The polydisperse extinction efficiency factor versus the effective size parameter for $m_1' = 1$, $m_2' = 1.4$, and $m_1'' = 0$, 0.02, and 0.05. Each set of three extinction curves is labeled by the corresponding value of $v_{\text{eff}}$. The vertical scale applies to the curve for $v_{\text{eff}} = 0$, the other curves being successively displaced upward by 2. The results are displayed with a size-parameter step size of $\Delta x_{\text{eff}} = 0.01$.

Owing to the absence of total internal reflection, all curves in Fig. 4 are devoid of MDRs, whereas the bottom black curve in Fig. 5 exhibits a pronounced ripple structure. Given the (exceedingly) small widths of the MDRs, it is not surprising that as small an effective variance as $v_{\text{eff}} = 0.01$ serves to completely eradicate the fine ripple structure in Fig. 5 even when the host medium is nonabsorbing. What is quite surprising, however, is that the same extremely small width of the size distribution completely suppresses the phenomenon of negative extinction in both Figs. 4 and 5. Furthermore, as the effective variance increases, all interference oscillations subside and ultimately disappear. Even though the first interference maximum is still visible in the upper extinction curves computed for $v_{\text{eff}} = 0.15$, the right-hand part of each curve exhibits a smooth trend toward the canonical asymptotic value $Q_{\text{ext}} = 2$ consistent with the prediction of the extinction paradox [1,27,28].

We must therefore conclude that although negative extinction is undoubtedly a real physical phenomenon, observing it in practice may be highly nontrivial and necessitates a special laboratory setting involving a single particle or a small sparse group of nearly monodisperse particles.

As explained in Ref. [22], the computation of size-averaged quantities for given $r_{\text{eff}}$ and $v_{\text{eff}}$ involves a quadrature summation of the results obtained for a (very) large number of discrete radii distributed from essentially zero to several times the effective radius. In the process of these computations, we have detected unexpected overflows occurring for sufficiently large $m_1''$ and $x_{\text{eff}}$. We have been able to trace the origin of these overflows to the computation of the Lorenz–Mie coefficients, as follows.

According to Eqs. (37) and (38) of Ref. [22], the denominators of the formulas for $a_n$ and $b_n$ contain the Hankel function of the first kind,

$$h_n^{(1)}(m_1 x) = j_n(m_1 x) + i y_n(m_1 x), \qquad (11)$$

and its derivative

$$[m_1 x h_n^{(1)}(m_1 x)]' = \frac{d[m_1 x h_n^{(1)}(m_1 x)]}{d(m_1 x)}, \qquad (12)$$

where $j_n(m_1 x)$ is the spherical Bessel function of the first kind and $y_n(m_1 x)$ is the spherical Bessel function of the second kind. In the case of a nonabsorbing host medium, both $j_n(m_1 x)$ and $y_n(m_1 x)$ are real-valued, and hence neither $h_n^{(1)}(m_1 x)$ nor $[m_1 x h_n^{(1)}(m_1 x)]'$ vanishes. In the case of an absorbing host, both $j_n(m_1 x)$ and $y_n(m_1 x)$ are complex-valued. As a consequence, if

$$y_n(m_1 x) = i j_n(m_1 x) \qquad (13)$$

and

$$[m_1 x y_n(m_1 x)]' = i [m_1 x j_n(m_1 x)]' \qquad (14)$$

to a very high numerical accuracy then both $a_n$ and $b_n$ can vanish.

We have noticed that in many cases of large $m_1''$

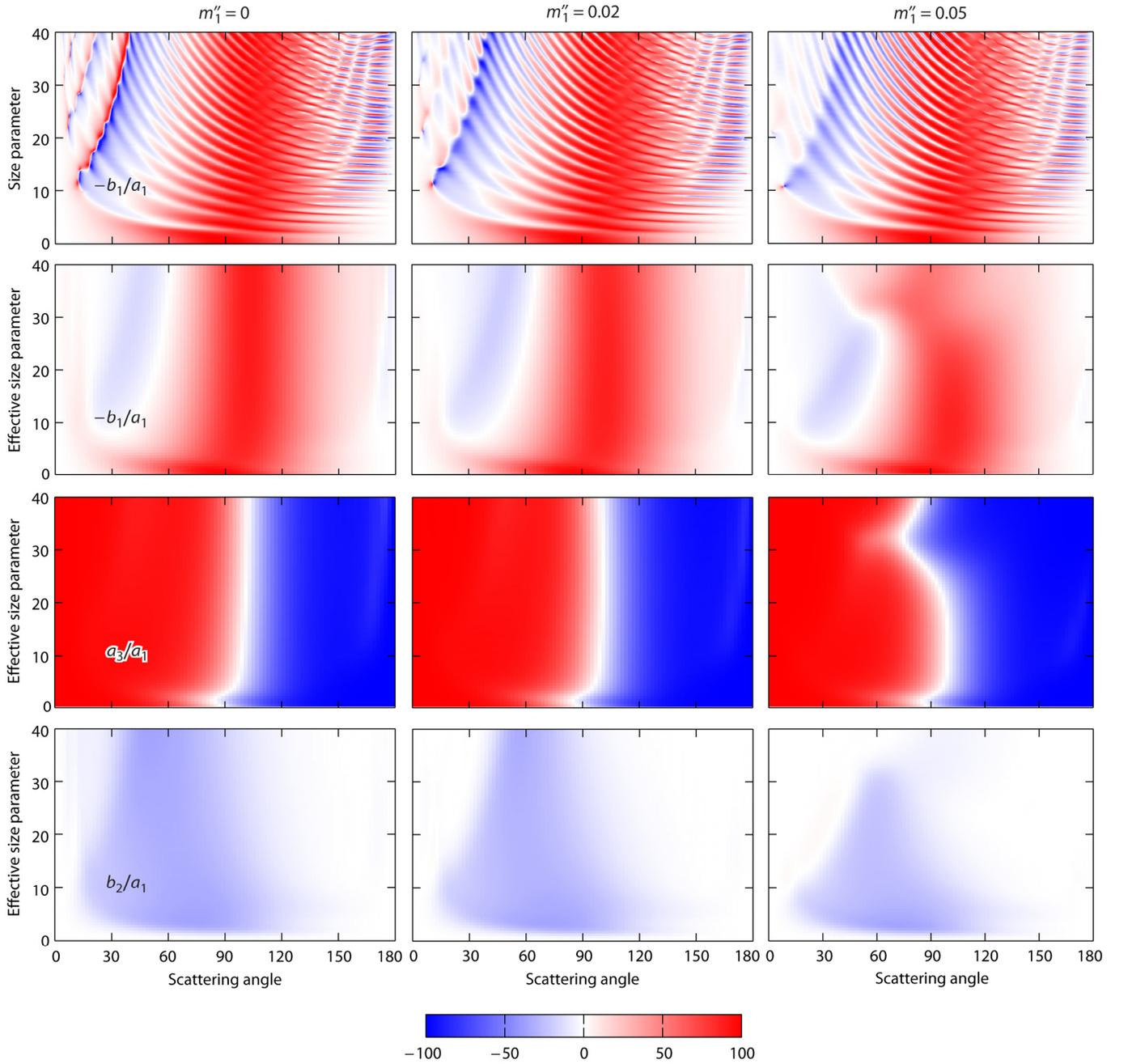

**Fig. 6.** Horizontal rows: ratios of select elements of the Stokes scattering matrix to the (1,1) element (in %) computed for $m_1' = 1.33$, $m_2' = 1$, and $m_1'' = 0$, 0.02, and 0.05. The vertical columns are labeled by the respective values of $m_1''$. The top row corresponds to $v_{\text{eff}} = 0$ and is plotted with scattering-angle and size-parameter resolutions $\Delta\Theta = 0.1°$ and $\Delta x = 0.1$, respectively. The three bottom rows correspond to $v_{\text{eff}} = 0.15$ and are plotted with scattering-angle and effective-size-parameter resolutions $\Delta\Theta = 2°$ and $\Delta x_{\text{eff}} = 0.5$, respectively.

---

and $x$ and certain $n$, the equalities (13) and (14) can hold despite the use of double-precision FORTRAN arithmetic, thereby leading to overflows. Yet it remains unclear whether the equalities (13) and (14) can hold in the strict mathematical sense or whether the use of extended-precision arithmetic can resolve this overflow issue in the numerical sense. Further analysis is obviously warranted.

**4. Stokes scattering matrix**

Another optical characteristic of great practical importance is the dimensionless so-called normalized Stokes scattering matrix describing the far-field transformation of the Stokes column vector of the incident plane wave into that of the scattered outgoing spherical wave provided that both columns are defined with respect to the scattering plane [22,35]. This matrix is given by

$$\tilde{\mathbf{F}}(\Theta) = \begin{bmatrix} a_1(\Theta) & b_1(\Theta) & 0 & 0 \\ b_1(\Theta) & a_1(\Theta) & 0 & 0 \\ 0 & 0 & a_3(\Theta) & b_2(\Theta) \\ 0 & 0 & -b_2(\Theta) & a_3(\Theta) \end{bmatrix}, \quad (15)$$

where $\Theta \in [0, 2\pi]$ is the scattering angle (i.e., the angle between the incidence and scattering directions). In the case of unpolarized incident light, the element $a_1(\Theta)$, called the phase function, describes the angular distribution of the scattered intensity, while the ratio $-b_1(\Theta)/a_1(\Theta)$ specifies the corresponding degree of linear polarization. The phase function is normalized according to

$$\frac{1}{2}\int_0^\pi d\Theta \sin\Theta\, a_1(\Theta) = 1. \quad (16)$$

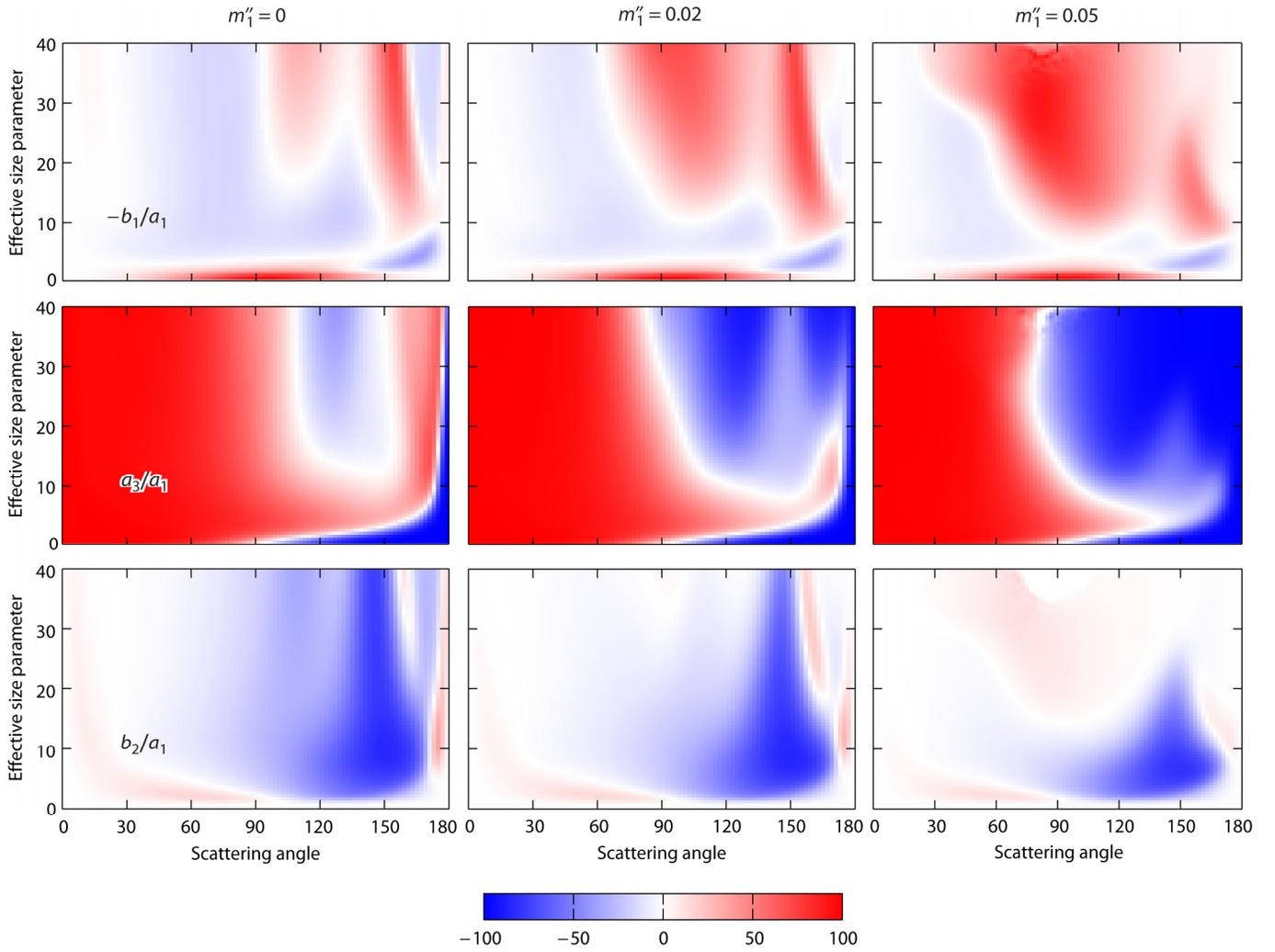

**Fig. 7.** Horizontal rows: ratios of select elements of the Stokes scattering matrix to the (1,1) element (in %) computed for $v_{\text{eff}} = 0.15$, $m'_1 = 1$, $m'_2 = 1.4$, and $m''_1 = 0$, 0.02, and 0.05. The vertical columns are labeled by the respective values of $m''_1$. The results are displayed with scattering-angle and effective-size-parameter resolutions $\Delta\Theta = 2°$ and $\Delta x_{\text{eff}} = 0.5$, respectively.

In general,

$$a_3(0) = a_1(0), \quad (17)$$

$$a_3(180°) = -a_1(180°), \quad (18)$$

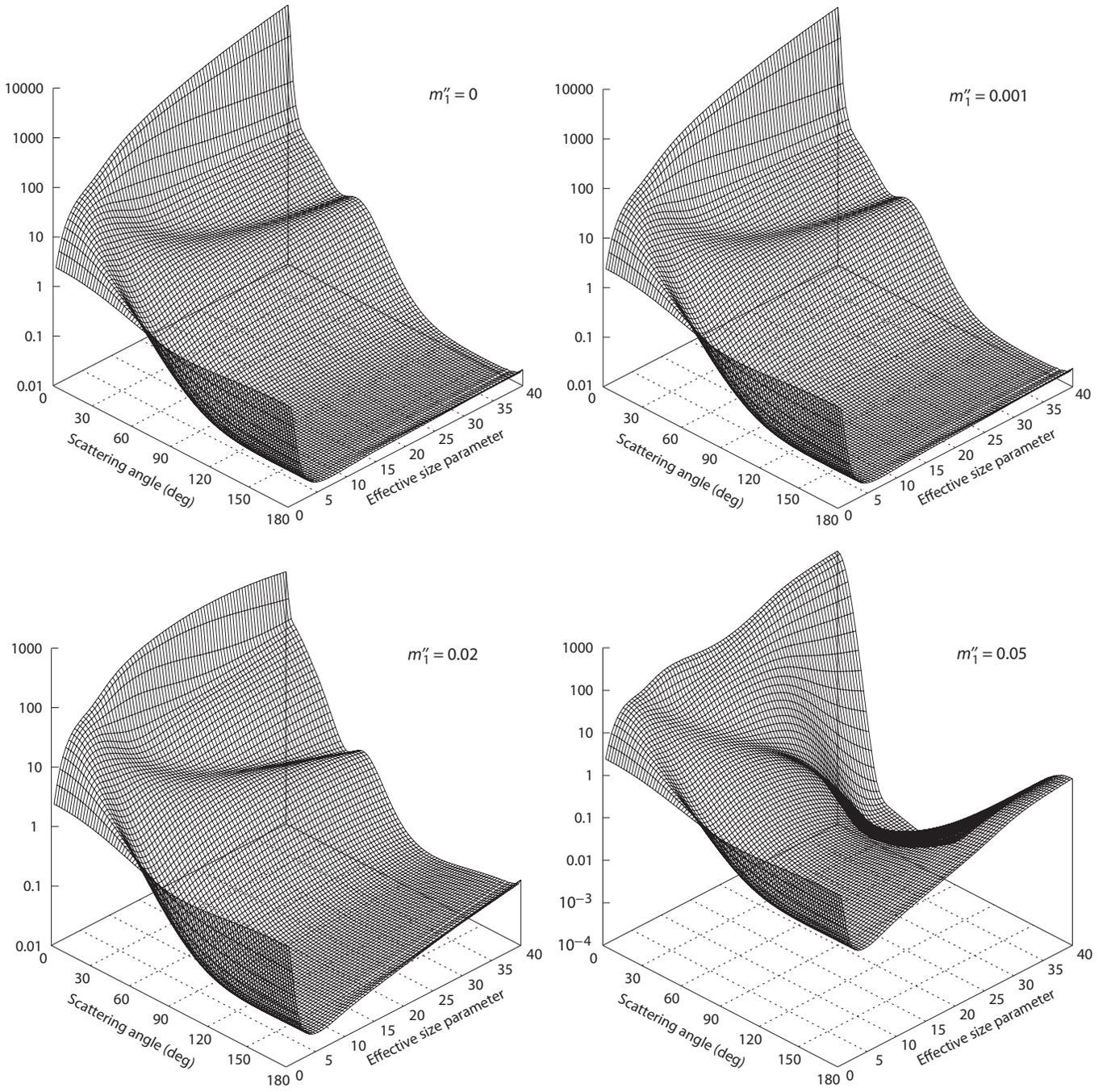

**Fig. 8.** The phase functions plotted for $m'_1 = 1.33$, $m'_2 = 1$, and $m''_1 = 0$, 0.001, 0.02, and 0.05 with scattering-angle and effective-size-parameter resolutions $\Delta\Theta = 2°$ and $\Delta x_{\text{eff}} = 0.5$, respectively.

and

$$b_1(0) = b_2(0) = b_1(180°) = b_2(180°) = 0. \qquad (19)$$

The upper left panel of Fig. 6 is plotted for the pair $\{m_1 = 1.33, m_2 = 1\}$ with high scattering-angle and size-parameter resolutions. It shows that in the case of monodisperse particles, adding the scattering angle as an independent variable parameter results in a very complex two-dimensional interference structure, with nearly −100% to nearly +100% polarization swings often occurring over extremely narrow scattering-angle and size-parameter intervals. Growing absorption in the host medium serves to increasingly suppress this structure. Averaging over the moderately wide gamma distribution of particle radii with an effective variance of $v_{\text{eff}} = 0.15$ (the second row of panels) eliminates the interference structure completely, thereby making the ratio $-b_1(\Theta)/a_1(\Theta)$ a rather smooth function of the scattering angle and effective size parameter. The same applies to the other two ratios of the scattering-matrix elements plotted in the bottom two rows of panels. In particular, increasing $m''_1$ serves to expand the domain of near-neutral values of the size-averaged ratio $b_2(\Theta)/a_1(\Theta)$. In the limit $x_{\text{eff}} \to 0$, all diagrams exhibit the typical features of Rayleigh scattering: the degree of linear polarization is symmetric with respect to $\Theta = 90°$ and tends to 100% at $\Theta = 90°$, the ratio $a_3(\Theta)/a_1(\Theta)$ becomes antisymmetric with respect to $\Theta = 90°$, and the ratio $b_2(\Theta)/a_1(\Theta)$ vanishes.

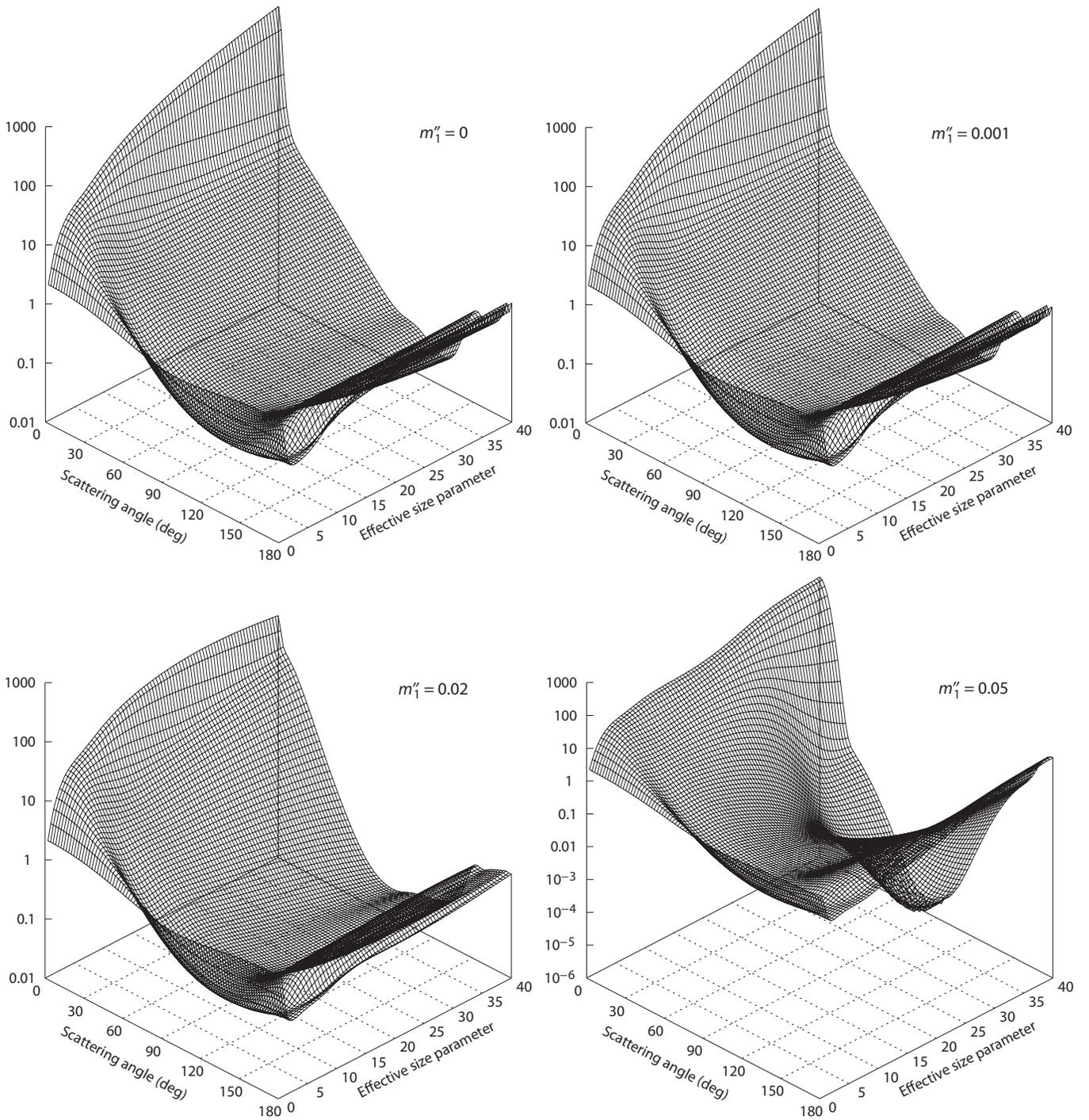

**Fig. 9.** The phase functions plotted for $m'_1 = 1$, $m'_2 = 1.4$, and $m''_1 = 0$, 0.001, 0.02, and 0.05 with scattering-angle and effective-size-parameter resolutions $\Delta\Theta = 2°$ and $\Delta x_{\text{eff}} = 0.5$, respectively.

Analogous plots for the pair $\{m_1' = 1, m_2' = 1.4\}$ (Fig. 7) exhibit a much more pronounced effect of increasing absorption in the host medium. In particular, the areas of positive polarization expand dramatically (the top row of panels); the areas of negative values of the ratio $a_3(\Theta)/a_1(\Theta)$ at side- and backscattering angles expand to encompass almost the entire backscattering hemisphere (the middle row of panels); and the ratio $b_2(\Theta)/a_1(\Theta)$ becomes substantially more neutral (the bottom row of panels).

Figures 8 and 9 parallel Figs. 6 and 7, respectively, and depict the corresponding phase functions. It is seen that the phase functions for $m_1'' = 0$ and 0.001 are virtually indistinguishable and that the limit $x_{\text{eff}} \to 0$ of Rayleigh scattering remains immune to increasing $m_1''$. As the effective size parameter increases, the phase functions in the upper two panels of Fig. 8 develop a nearly horizontal "shelf" at scattering angles around $\Theta \approx 45°$. As explained on page 264 of Ref. [35], this feature has a geometric-optics origin and is typical of situations wherein $m_2' < m_1'$. It is seen that increasing $m_1''$ effectively destroys this shelf of phase-function values. As $m_1''$ reaches 0.05, the phase functions at larger effective size parameters develop a very deep minimum at side-scattering angles bracketed by a strong diffraction peak in the forward direction and a relatively strong backscattering maximum. This effect is especially remarkable in the case of $m_1' = 1.33$ and $m_2' = 1$ since outside the narrow Rayleigh region of effective size parameters, the corresponding phase functions computed for $m_1'' = 0$, 0.001, and 0.02 are almost completely featureless at backscattering and near-backscattering angles. The side-scattering minimum is especially deep in the case of $m_1' = 1$, $m_2' = 1.4$, and $m_1'' = 0.05$, with phase-function values dropping well below $10^{-5}$.

## 5. Concluding remarks

The main objective of this paper has been to apply the FORTRAN program developed in Ref. [22] to a limited yet representative set of problems and thereby extend the analysis performed in Chapter 9 of Ref. [35] to the case of an absorbing host medium. Needless to say, although this extension has already yielded several interesting and unexpected results (such as the phenomenon of negative extinction and its complete disappearance upon averaging over a very narrow size distribution), further research and applications to specific practical problems are needed. The range of $\{m_1, m_2\}$ combinations encountered in natural and artificial environments is quite wide, and it is imperative to investigate the effects of realistic rather than model values of absorption in the host medium. Another natural extension of our analysis is to apply the generalized first-principles radiative transfer theory developed in Refs. [36,37] to large sparse multi-particle groups (i.e., clouds of particles) embedded in a lossy host medium.

## Acknowledgments

We appreciate numerous insightful discussions with James Lock, Gorden Videen, and Ping Yang and thank Nadezhda Zakharova for help with graphics. This research was supported by the NASA Radiation Sciences Program (manager Hal Maring) and the NASA Remote Sensing Theory Program (manager Lucia Tsaoussi). MIM thanks the organizers of the APOLO-2017 conference Oleg Dubovik and Zhengqiang Li for travel support.